\documentclass[conference]{IEEEtran}
\IEEEoverridecommandlockouts
\usepackage{cite}
\usepackage{amsmath,amssymb,amsfonts}
\usepackage{algorithmic}
\usepackage{graphicx}
\usepackage{textcomp}
\usepackage{xcolor}
\def\BibTeX{{\rm B\kern-.05em{\sc i\kern-.025em b}\kern-.08em
T\kern-.1667em\lower.7ex\hbox{E}\kern-.125emX}}
\begin{document}

\title{A Many-ported and Shared Memory Architecture for High-Performance ADAS SoCs\\

}
\author{\IEEEauthorblockN{Hao Luan}
\IEEEauthorblockA{\textit{Horizon Robotics } \\
}
\and
\IEEEauthorblockN{Yu Yao}
\IEEEauthorblockA{\textit{Horizon Robotics } \\
}
\and
\IEEEauthorblockN{Chang Huang}
\IEEEauthorblockA{\textit{Horizon Robotics } \\
}
\thanks{The pre-print version of the same paper that will be presented in NoCs 2022.}

}

\maketitle

\begin{abstract}
Increasing investment in computing technologies and the advancements in silicon technology has fueled rapid growth in advanced driver assistance systems (ADAS) and corresponding SoC developments. \\
An ADAS SoC represents a heterogeneous architecture that consists of CPUs, GPUs and artificial intelligence (AI) accelerators. In order to guarantee its safety and reliability, it must process massive amount of raw data collected from multiple redundant sources such as high-definition video cameras, Radars, and Lidars to recognize objects correctly and to make the right decisions promptly. A domain specific memory architecture is essential to achieve the above goals. \\
We present a shared memory architecture that enables high data throughput among multiple parallel accesses native to the ADAS applications. It also provides deterministic access latency with proper isolation under the stringent real-time QoS constraints. A prototype is built and analyzed. The results validate that the proposed architecture provides close to 100\% throughput for both read and write accesses generated simultaneously by many accessing masters with full injection rate. It can also provide consistent QoS to the domain specific payloads while enabling the scalability and modularity of the design.\\

\end{abstract}


\section{Introduction}\label{sec-Intro}
Autonomous driving systems have attracted significant interest recently. Many existing industry leaders, such as Tesla\cite{b2}, Nvidia\cite{b5}, Mobileye and Huawei\cite{b4}, and a few startups such as Horizon Robotics, have invested a large amount of capital and engineering power on developing ADAS SoCs. \\

There are six levels of automation defined by SAE International\cite{b12}, level 1 and 2 of automation are mostly
driving assistance, where the human driver still handles a
substantial portion of the driving tasks under \textit{all} conditions. Autonomous driving systems can take full
driving responsibility at level 3-5 of automation under \textit{certain}
driving conditions, which are typically referred to as highly autonomous vehicles 
(HAVs). Because HAVs represent the future of autonomous driving systems, we focus on HAVs at level 3-5 for the rest of this paper. \\

A typical SoC serves for HAVs consists of the following processing pipelines and it should be able to complete it under current traffic conditions with a latency of 100 ms at a frequency of at least once every 100 ms or less\cite{b1}: 
\begin{enumerate}
\item \textbf{Object identification and tracking}: The video captured by cameras or other formats of raw data, such as data sourced by Light Detection and Ranging (LIDAR) and Radar\cite{b3}, are streamed into both the object detection engine to detect objects and the localization engine to locate the vehicle in parallel. This step of processing is often accomplished with various machine-learning (ML) algorithms assisted by CPUs and other types of acceleration such as image cropping and Region of Interest (ROI) identification. Massive data is collected real-time and a large number of training data sets (thousands of images) are also needed to infer and differentiate among vehicles, common objects such as traffic signs, pedestrians, and streets etc.
\item \textbf{Fusion}: The vehicle location and tracked objects are projected into the same 3D-coordinate space by the fusion engine, which is a processing step that combines the prior results into an understandable result. Again, a lot of heterogeneous processing occurs in this step, and there is a lot of data sharing between this step and the previous one.
\item \textbf{Motion Planner}: In this step, the processed and analyzed data is now consumed by the motion planner to make operation decisions. The vehicle's moving track and projected paths are calculated and information such as navigation data is correlated. Heavy computations with a smaller and condensed data set is the characteristic of this processing step compared to the other steps.
\item \textbf{Control and Decision}: The last and most important step in the entire flow. This step normally has the most stringent safety requirements to make sure the action and moving path are thoroughly checked and verified with the redundant pool of computing resources. Reliability is the key characteristic; however, it has the least amount of data to handle and process. 
\end{enumerate} 

To fulfill such a real-time pipeline, one can easily observe that abundant ML and other heterogenous processing power is undoubtably needed on silicon. An efficient and domain specific shared memory architecture with the following characteristics is also critical to keep all of them realizing at high TOPS (Tera Operations Per Second) computing capacity:
\begin{itemize}
\item Big in size: DaDianNao\cite{b9}, a ML supercomputer, employs 36MB on-chip shared memory to accelerate machine learning applications. Tesla’s full self-driving
(FSD) SoC uses two instances of neural-network accelerator (NNA), where each NNA has a 32 MB of on-chip SRAM memory providing a high throughput data feed to enable and sustain a total of 72 TOPS processing power\cite{b2}. Huawei Ascend 901 SoC has two instances of 16 MB on-chip shared memory to enable a 256 TOPS processing power\cite{b4}. Therefore, a size of 32 MB of shared memory can be assumed as the baseline to accelerate heterogenous ADAS processing with a lot of ML involved.
\item It is a many-ported and a parallel architecture: The data is shared among many relevant heterogeneous processing elements. It is always being accessed in parallel and to provide high data throughput to all of them.
\item Provides consistent access latency optimized for the HAVs payloads\cite{b11}: the raw image data, LIDAR, Radar data and the model data for ML processing are always available in the range of KB or MB. Thus, the architecture needs to fully utilize the bulky access nature of image processing and neural networks from a buffer level access as opposed to a single random byte or word access.
\item Provides necessary isolation: special care must be taken to provide sufficient isolation among the multiple parallel data paths to comply ISO 26262\cite{b7} requirements.
\item Software friendly: it should not behave as yet another conventional discrete memory on an SoC, where different pieces of the memory manifests different access latency and QoS. It needs to present to the software programmer as a flat and uniform memory space with consistent access latency across the entire memory space. This requirement can significantly ease the efforts of software programming and code management.
\item Scalable: the same architecture can easily migrate
from generation to generation from both logical and physical design perspectives and across multiple silicon manufacturing nodes.
\end{itemize}

The rest of the paper is organized as follows: Section \ref{sec-Related-and-Arch} analyzes the architectural challenges brought by the above requirements, surveys
related work, presents our architecture, and describes the key technologies in detail. In Section \ref{sec-implementation-Res}, we share the implementation details and relevant results
to validate the effectiveness of our approach, while Section \ref{sec-Conclusion} summarizes
our conclusions and findings.

\section{Related Work and Architecture}\label{sec-Related-and-Arch}
\subsection{Related Work}\label{subsec-Related-work}
SoC architectures based on the shared memory are the preferred backbone for flexible and programmable solutions in many application domains. Many-ported shared memory architecture has been explored recently both in academic and industrial settings. \\

Mesh-of-Trees (MoT) topology has shown that it can provide a consistent throughput as high as 98\% for up to 64 masters and memory modules\cite{b10}. MoT topology consists of two phases: the routing phase and the arbitration phase. In the routing phase, every level of fan-out routing trees dilutes the traffic conflicts by half. This splitting process repeats as many as $log_2 N$ hops, where $N$ is the number of masters. Hereafter, the data paths are merged by two gradually in the arbitration phase, which is also repeated in $log_2 N$ hops and the total data points are converge back to $N$ number of memory modules. The separation and isolation is well maintained in the routing phase. The splitting process is also an effective scheme to mediate memory access contentions. However, the isolation among multiple accessing masters, which is highly desirable in HAVs, is gradually lost in the arbitration phase. Besides, the architecture dictates a flat structure. This makes it harder to support modularity and scalability, which is very much required in the engineering practice. \\

DaDianNao\cite{b9}, which is a ML supercomputer, uses a Fat Tree topology consisting of total 16 data tiles to provide a 36 MB shared memory at a 28 nm technology node. It is big in capacity; however, the wires of the interconnect occupy half of the die area of each tile due to congestion. The two-level hierarchical structure is good to scale up the design but the global resource sharing has an intrinsic NUMA nature, depending on whether the producer and consumer of the data are on the same level or not. \\

Kalray MPPA-256\cite{b8}, a manycore processor targets real-time and embedded applications including ADAS. It has a total of sixteen homogenous clusters, and each cluster has a 2 MB shared memory, which is shared by 17 identical VLIW (Very Long Instruction Word) cores without cache coherency. The shared memory comprises of 16 independent memory banks. The memory banks are arranged in two sides (left and right). The PEs are organized in 8 pairs. Each pair has two memory buses (one for each side of memory group), which can be utilized in parallel by the two cores. As illustrated in \cite{b8}, the interconnect topology employs two levels of crossbar switches where the first arbitration is done between the two cores in a pair, then the next level of arbitration is done among all pairs. The topology reduces a full crossbar of 68 x 16 to two separate full crossbars of 17 x 8. It is effective to reduce the overall number of wires, arbitration costs and implementation complexity, and to enable a shared memory of 2 MB with uniform memory access using a TSMC 28 nm HP process. However, the multi-level crossbars and round robin arbitrations really throttle the overall throughput, and the software/system engineers need to figure out how to avoid the memory access conflicts among all the cores. Secondly, the homogeneity of the PE and its physical shape enables a perfect layout where all the PEs are sandwiched between the memory banks located on the both sides\cite{b8}. This is something rarely available on an SoC consists of many heterogenous PEs along the data path.\\

A distributed and modular architecture is presented and studied in\cite{b6} to mainly optimize the real-time payloads for the wireless applications while it tries to reduce the interconnect area at the same time. The architecture supports 32 masters with a size of 16MB of shared memory, which is implemented on a 16nm technology node. The architecture employs multi-level low index switches to further reduce the wire complexity and arbitration costs that natively exists in many-ported interconnect and manages to confine the interconnect area to be less than 30\% of the total area. The hierarchical and modular approach is effective to make the overall architecture more scalable and easier to implement. The distributed approach can help to mediate memory access contentions and is also effective to mediate the Non-Uniform Memory Access (NUMA) effects. However, the divide and conquer approach taken to optimize interconnect area cannot satisfy the isolation requirement since it has merged the traffic from multiple parallel accessing masters right at the beginning of the data path. 
\subsection{Architectural Challenges and Considerations}\label{subsec-Tech-chan}
Based on the analyses of the available architectures, the interconnect and its topology is the key factor that drives overall performance for a many-ported shared memory with a big capacity. To meet the ADAS requirements outlined in section \ref{sec-Intro}, important parameters such as how to connect many master ports to the sea of the memory instances, how to route, and to arbitrate the traffic generated by the parallel accessing masters need to be considered. Here is the summary of the challenges:\\
%\begin{itemize}
\begin{enumerate}
\item The size matters: A shared memory with a size of 32 MB or above normally consists of at least half a thousand to multiple thousand pieces of physical memory instances depending on the density and isolation requirements. The physical layout is very memory dominant and the Non-Uniform Memory Access (NUMA) effect is even worse with the growth of the size and the number of memory instances. Therefore, the physical implementation difficulties should be an important factor to consider.
\item Mediation of the memory access contention: Sharing by nature brings in access contentions among multiple parallel accesses generated by many heterogenous processing elements. The interconnect, memory grouping and memory address mapping all need to be considered together to reach an efficient and cost-effective architecture.
\item Isolation and sharing are a pair of contradicting requirements: The contents of the memory need to be shared as much as possible to keep it local, and reduce the power of data movement. However, the accessing data paths from the accessing masters also need to be kept isolated as much as possible to avoid unnecessary interference, which may be required after the ASIL allocation pursuant to ISO 26262\cite{b7}. 
%\end{itemize}
\end{enumerate} 
\subsection{The Domain Specific Memory Architecture }\label{subsec-Arch-work}
\begin{figure}[htbp]
\centerline{\includegraphics[width=\linewidth, height=8cm]{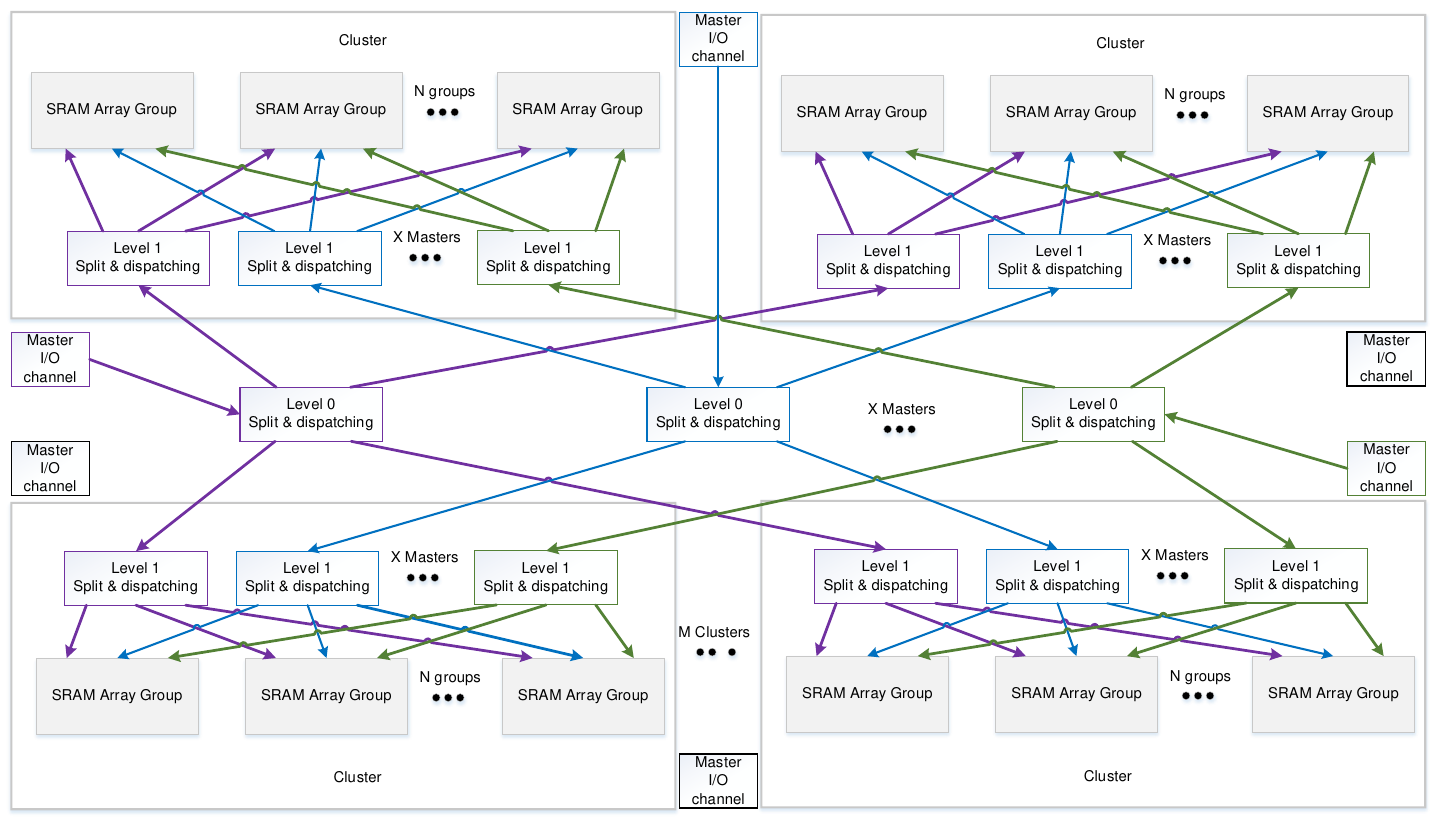}}
\caption{The Architectural View of the Proposed Shared Memory Controller}
\label{arch-figure}
\end{figure}

\begin{figure}
\includegraphics[width=0.7\linewidth,height=0.6\linewidth]{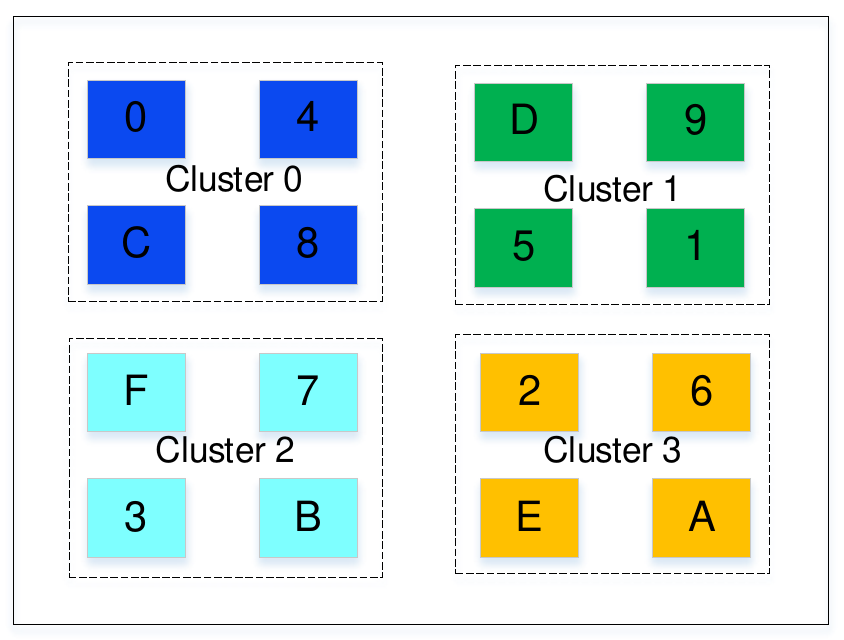}
\centering
\caption{Structural and Fractal Randomization Applied at Inter and Intra Cluster Level }
\centering
\label{Randomization-BB} 
\end{figure}

Fig. \ref{arch-figure} shows the overall architecture. It is a distributed and hierarchical one, which is applicable from both logical and physical design perspective. A symmetric logical and physical partition is employed to increase the scalability and ease of implementation. The interconnect between the masters and the sea of memory instances is comprehended by a multiple level recursive splitting and distributing structure. Just for simplicity, a two-level split and dispatching structure is shown to explain the concepts of the architecture. The bigger the size of shared memory, the more levels can be used to increase the parallelism. Therefore, less memory access contentions can be achieved along with more parallelism enabled by this multi-level splitting process. \\
\\
The splitting and dispatching is carried out in a recursive manner. Split by four, eight or even sixteen can be considered based on the shortest burst size among most frequently used burst sizes. For example, if burst four, eight and sixteen are the frequently used burst sizes on an SoC, A split-by-four structure is recommended. A recursive split-by-four architecture is shown in Fig. \ref{Randomization-BB} to explain the splitting and dispatching scheme due to the popularity of this combination on today's SoCs. \\
The following summarizes the rules to split and distribute a multi-beat read command and write data inside the proposed architecture: 
\begin{enumerate}
\item Disassemble any multi-beat read requests and write data, then spread them across four clusters once they enter the shared memory
\item Further introduce the next level of randomization so the multiple beats within a linear access go to a different SRAM array, to make sure it lands in a different memory bank to avoid access conflict inside a cluster and an SRAM array
\end{enumerate}

\begin{figure}
\centerline{\includegraphics[width=\linewidth, height=6cm]{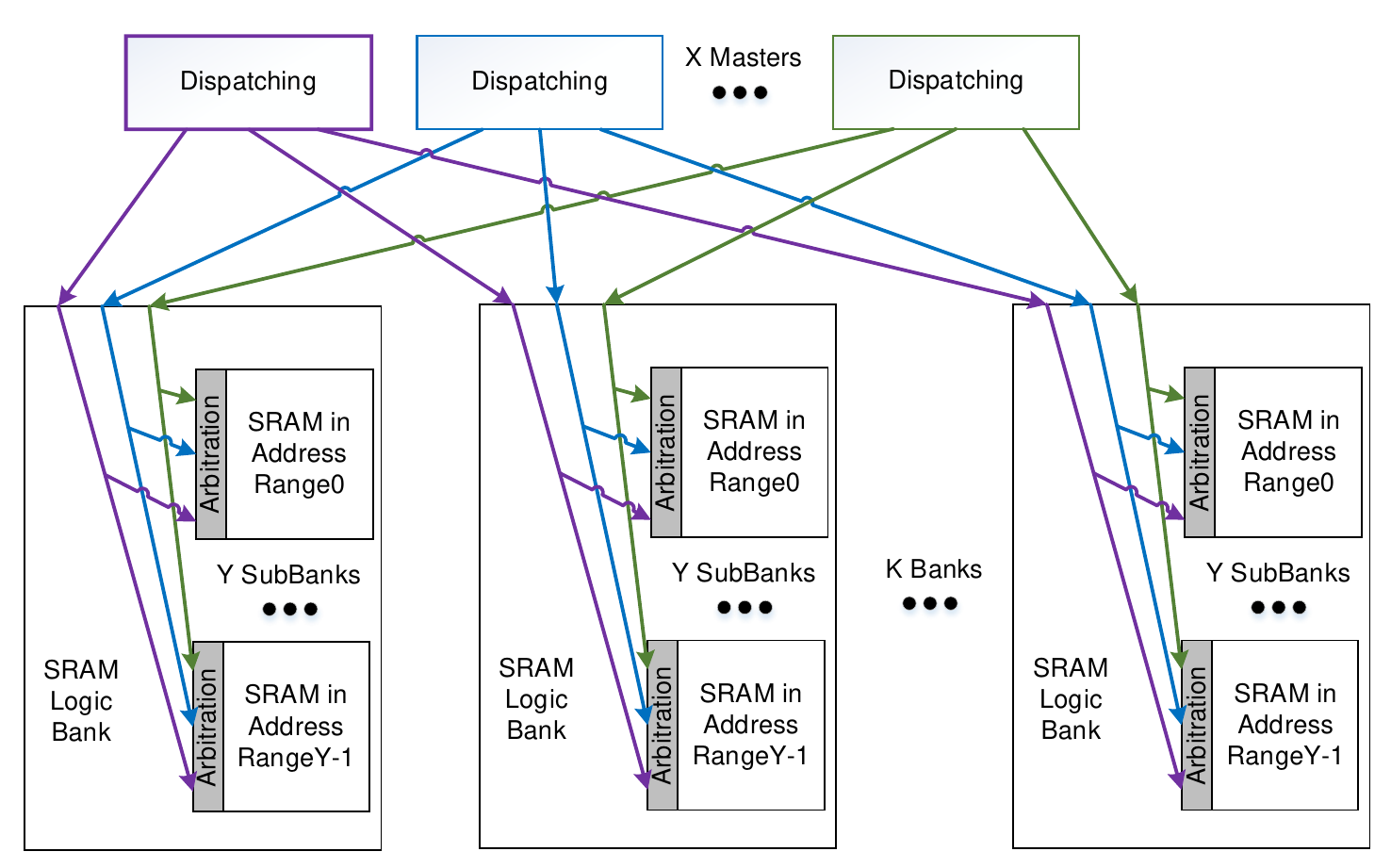}}
\caption{The Micro-Architecture of the SRAM Array}
\label{bank-figure}
\end{figure}

For example, if a burst four read command or write data is issued to the shared memory, every cluster gets one of the write data or read command beat as indicated by 0, 1, 2, 3 shown in different color. If a burst eight or sixteen read command or write data is issued to the shared memory, each cluster gets two or four beats of the read command or write data as shown in Fig. \ref{Randomization-BB} and they are guaranteed to fall in different SRAM arrays by applying the rules above. Enough randomization and whitening effects are introduced by applying the rules; the NUMA effect can be mediated significantly by averaging the individual beat access within a burst to achieve a consistent access latency to all multi-beat read and write accesses. The above mechanisms successfully address architectural challenge No.1 and No.2.\\
The micro-architecture of an SRAM array is shown in Fig.\ref{bank-figure}. At the input of an SRAM array, all the single or multi-beat burst transactions have been split into single beat transactions in the previous split and dispatching structure, matching the data width of the SRAM instances inside. The dispatching logic decodes and routes the beat transactions to $K$ logic banks based on the programmed addressing scheme. The arrangement of the memory instances is the direct reflection of the address scheme, which is the area where the grouping of the sea of the memory can be organized in a way to address the multiple requirements of ADAS applications. Schemes such as interleaving or hashing etc., are helpful to increase the access parallelism of the logic banks when multiple masters try to access the same address range at the same time. The memory instances are grouped in a two-dimensional manner with a goal to further increase the parallelism to reduce the access collisions while maintain access isolation to some extent. Within an SRAM logic bank, the memory instances are sliced by regions. The two-dimensional arrangement completes a linear memory space by many vertical layers where each layer is realized by the sub-banks fall in the same region across multiple memory banks. \\

For each logic bank, depending on the function safety and isolation requirement, the SRAM instances are separated into different sub-banks where each sub-bank has its independent arbitration logic. Together with each master’s independent data path before the arbitration, this architecture can provide complete separation for two masters accessing different sub-banks. Please note if there isn't any isolation required among multiple regions inside a memory logic bank, one arbitration logic should be sufficient for the whole bank. \\

With the replication of the arbitration logic, the data path from one accessing master to a memory region can be totally isolated with that of another accessing master accessing a different memory region. This scheme addresses challenge No.3 and provides the fundamental support to satisfy the ASIL-B requirement defined by ISO 26262 \cite{b7} along with other safety mechanisms such as ECC (Error Correcting Code) and time-out on the command and data. One may achieve ASIL-D results by leveraging system and software algorithms to compare and validate two ASIL-B data from two independent sources together\cite{b7}. The number of the logic banks in an SRAM array, how many sub-banks in a logic bank and the number of the accessing masters supported are cost sensitive to the total area and power consumption of the architecture. Therefore, these three important parameters need to be examined properly to achieve a good function/cost ratio. 
\section{Implementation, Results and Correlations with the Architecture}\label{sec-implementation-Res}
Multiple configurations have been studied in this architectural exploration. One prototype with the following configurations has been validated end-to-end from RTL simulations all the way to the physical design sign-off.
\begin{enumerate}
\item A shared memory with a size of 32 MB consists of over half an thousand instances of physical memory 
\item Sixteen accessing masters with 256 bit read and write data width employing AXI 5 protocol with read data chunking supported 
\item Two levels of splitting and distributing by four structures are employed to complete the interconnect of the shared memory: 
\begin{itemize}
\item Sixteen memory banks inside each SRAM array 
\item The shared memory has an outstanding capability of 8 commands right at every master port interface and an extra buffer worth of storing 64 splitting and dispatching beats
\item The interconnect network, which is represented by the two levels of splitting and distributing structure, runs at 1GHz
\item All memory instances run at 500 MHz
\end{itemize}
\end{enumerate}
To correlate with the overall architecture shown in Fig. \ref{arch-figure}, the above configuration can be summarized as: $X=16, M = 4, N=4 $, and each cluster has 8 MB of physical memory. 
\subsection{Simulation Results}\label{RTL-sims}
\begin{figure}[htb!]
%\centerline{\includegraphics[width=\linewidth, height=5cm]{table-read-No-masters-0509}}
\centerline{\includegraphics[width=\linewidth, height=5cm]{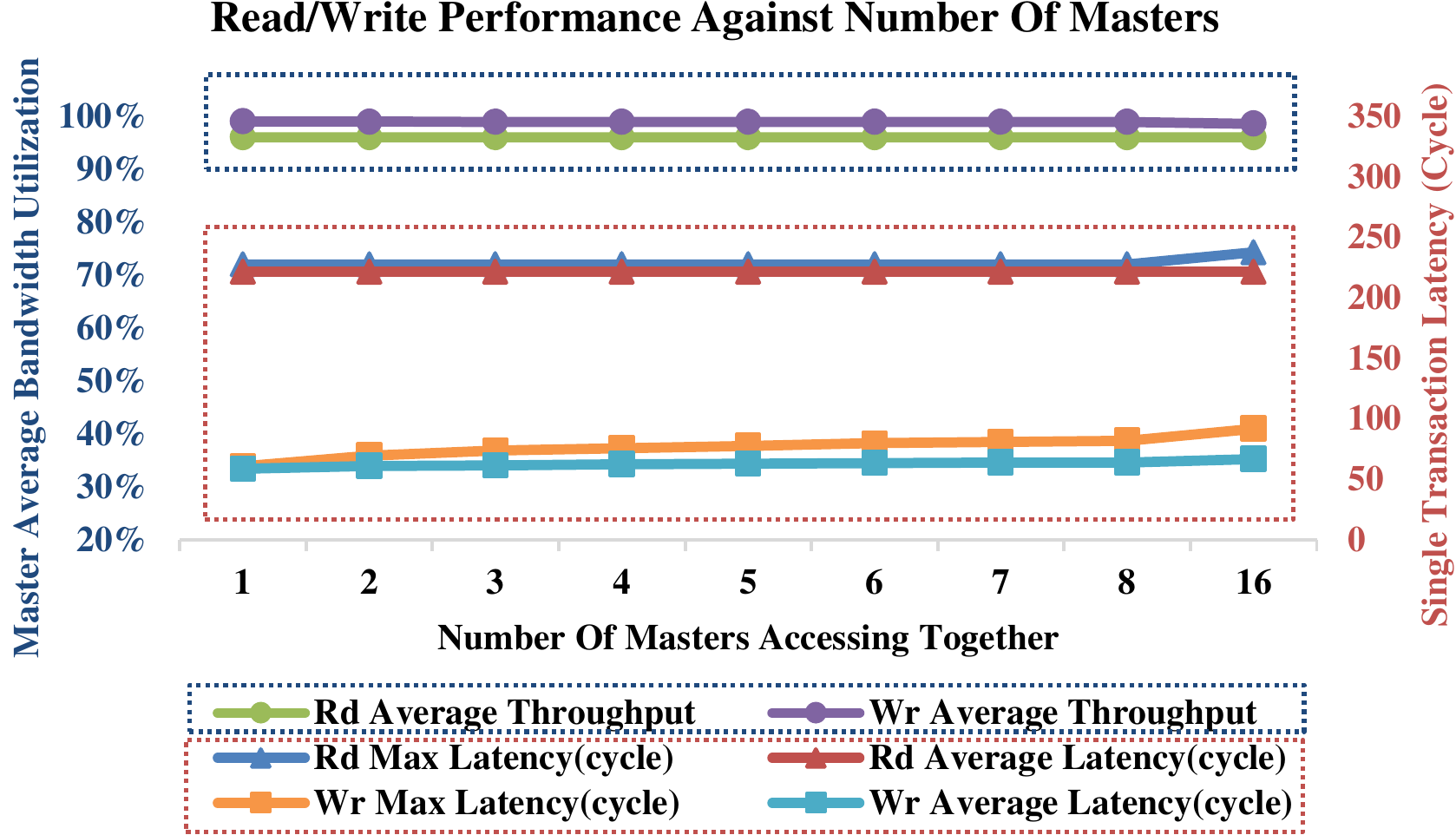}}
\caption{The Read and Write Performance with Different Number of Masters}
\label{sim-rd-wr-thput}
\end{figure}

In order to fully examine and validate the effectiveness of the proposed architecture, traffic is injected together using various number of the parallel accessing ports combined with different traffic patterns per accessing port. Each port is injected with $10,000$ read or write transactions in the simulation window. The throughput per port, the average read and write access latency for burst transfer and bulk data transfer performance are analyzed. Fig. \ref{sim-rd-wr-thput} shows the read and write performance with different numbers of parallel accessing masters ranging from one to sixteen. Each of the accessing master issues random read and write requests with random address (256bit aligned) at the same time with a 100\% injection rate. \\

Fig. \ref{sim-rd-wr-thput} clearly shows that even if the accessing master numbers vary from one to sixteen, the throughput per accessing port is stable for both the read and write transfers. The actual port throughput is around 96\% for the read traffic and is around 99\% for the write traffic. The port throughput does drop along with the addition of the accessing masters. However, the proposed architecture demonstrates a strong resilience to sustain the heavier traffic and only drops about $0.01$ percentage point for the read and about $0.46$ percentage point for the write throughout the whole range. This result is equal or better than that of MoT \cite{b10} topology but with a much coarser level of splitting and distributing structure, which directly translates to the cost saving of the implementation while keeping the scalability and modularity. The average read and write latency show the similar robustness on the same settings. Even though the maximum access latency degrading when more accessing masters are added, the average read latency stays almost the same, and the average write latency degrades just a few cycles. It validates that the multiple levels of splitting and distributing structure have fully randomized the read and write traffic so the NUMA effect can be tamed properly to provide consistent QoS. The results are better than those of \cite{b6} with a much simpler architecture. \\

Similar results can be obtained if a traffic consists of random burst four transactions, burst eight and even a combined traffic with three different burst lengths. Only the results based on burst sixteen traffic are shown just to avoid repeating the similar results. \\

\begin{figure}[htb!]
%\centerline{\includegraphics[width=\linewidth, height=5cm]{table-write-No-masters-0509}}
\centerline{\includegraphics[width=\linewidth, height=5cm]{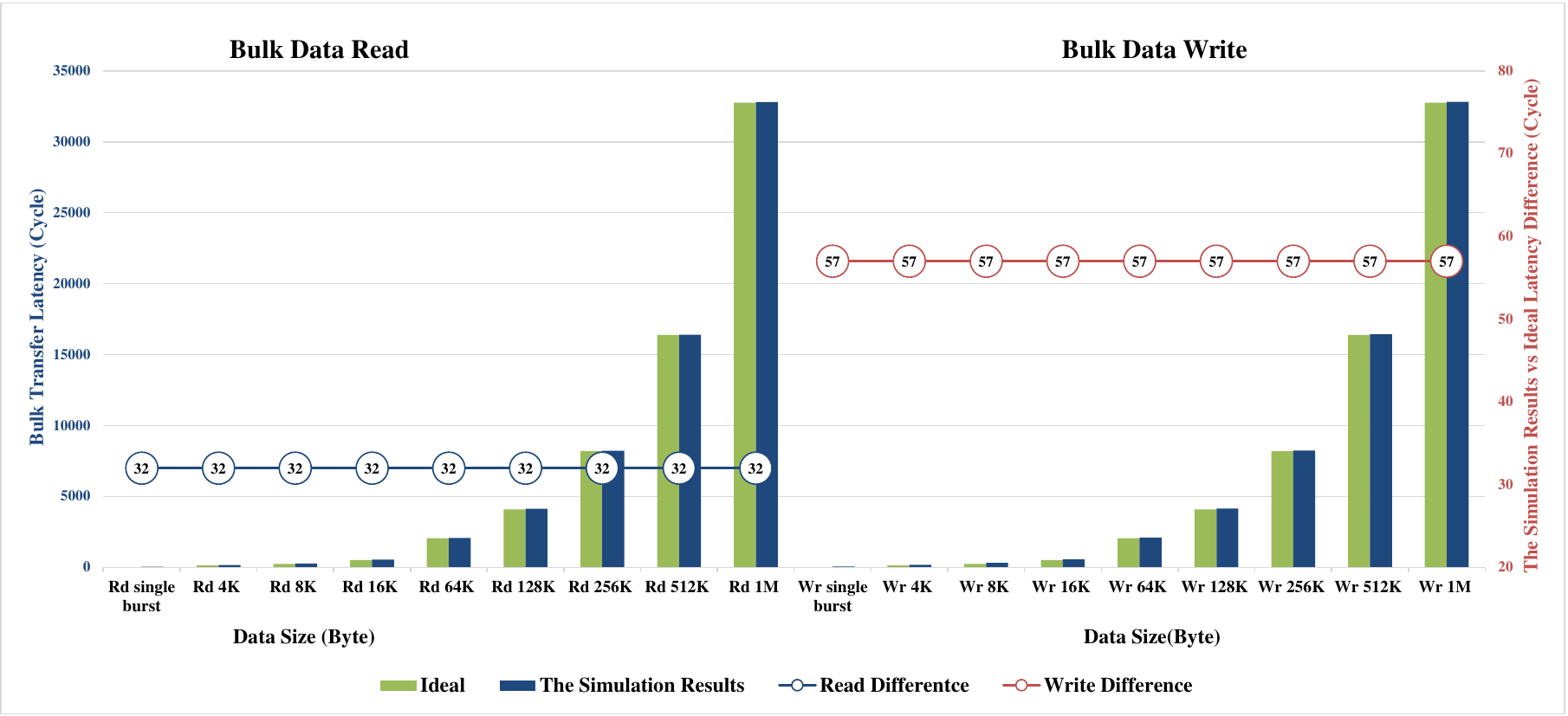}}
\caption{The Read and Write Latencies with Different Size of Payloads}
\label{sim-rd-wr-latency}
\end{figure}
To better mimic the real traffic patterns on an ADAS SoC, the second simulation is created to test the performance of bulk data transfers. This simulation is done with sixteen masters reading and writing bulk data at the same time from each and every accessing master in parallel. And all accessing memory spaces don't have any overlap to comply the isolation requirement. \\

\begin{figure}[htb!]
%\centerline{\includegraphics[width=\linewidth, height=6cm]{table-bulk-read-0509}}
\centerline{\includegraphics[width=0.95\linewidth, height=5cm]{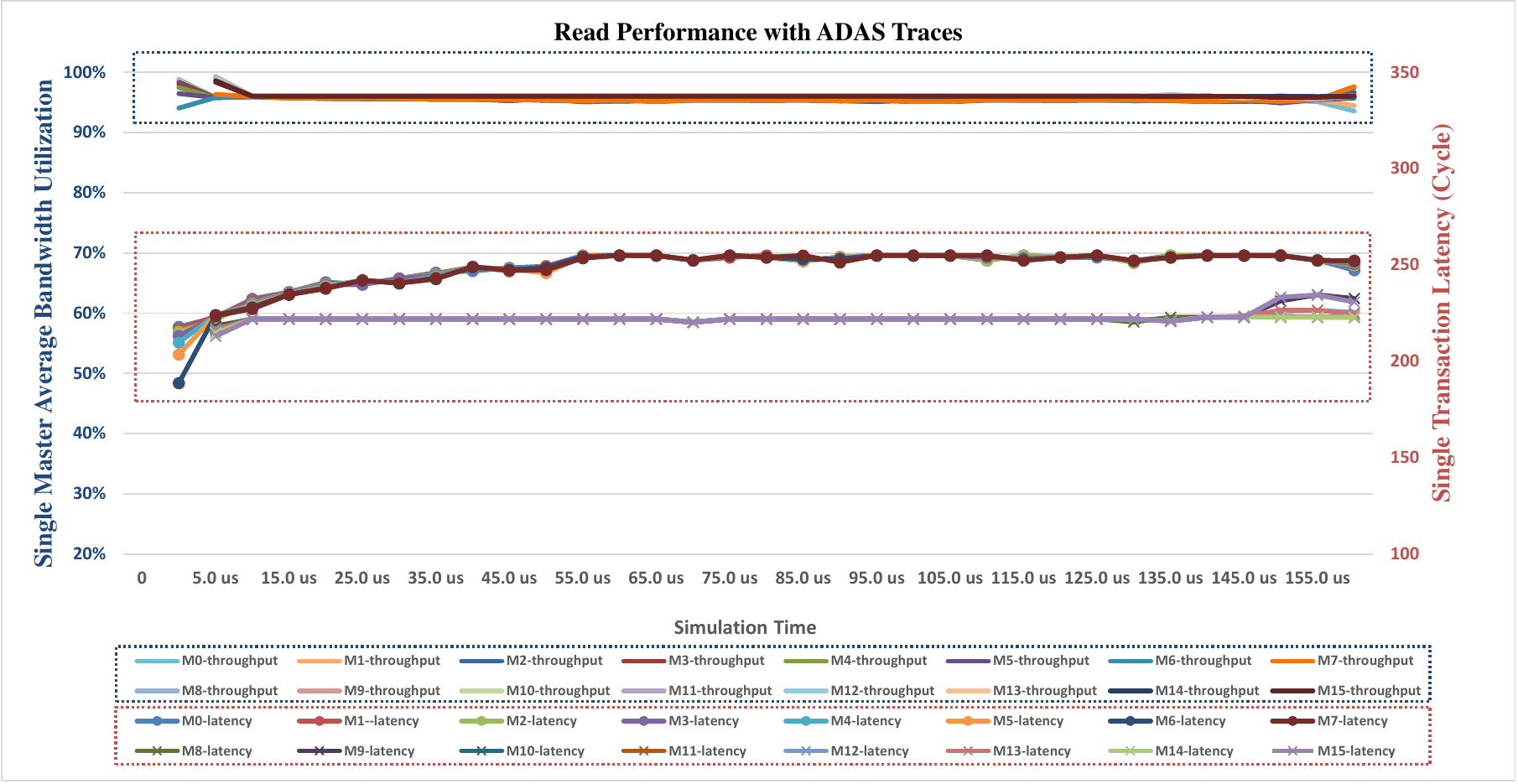}}
\caption{The Read Performance with Different ML and ADAS Traces}
\label{sim-rd-latency}
\end{figure}

\begin{figure}
%\centerline{\includegraphics[width=\linewidth, height=6cm]{table-bulk-write-0509}}
\centerline{\includegraphics[width=0.95\linewidth, height=5cm]{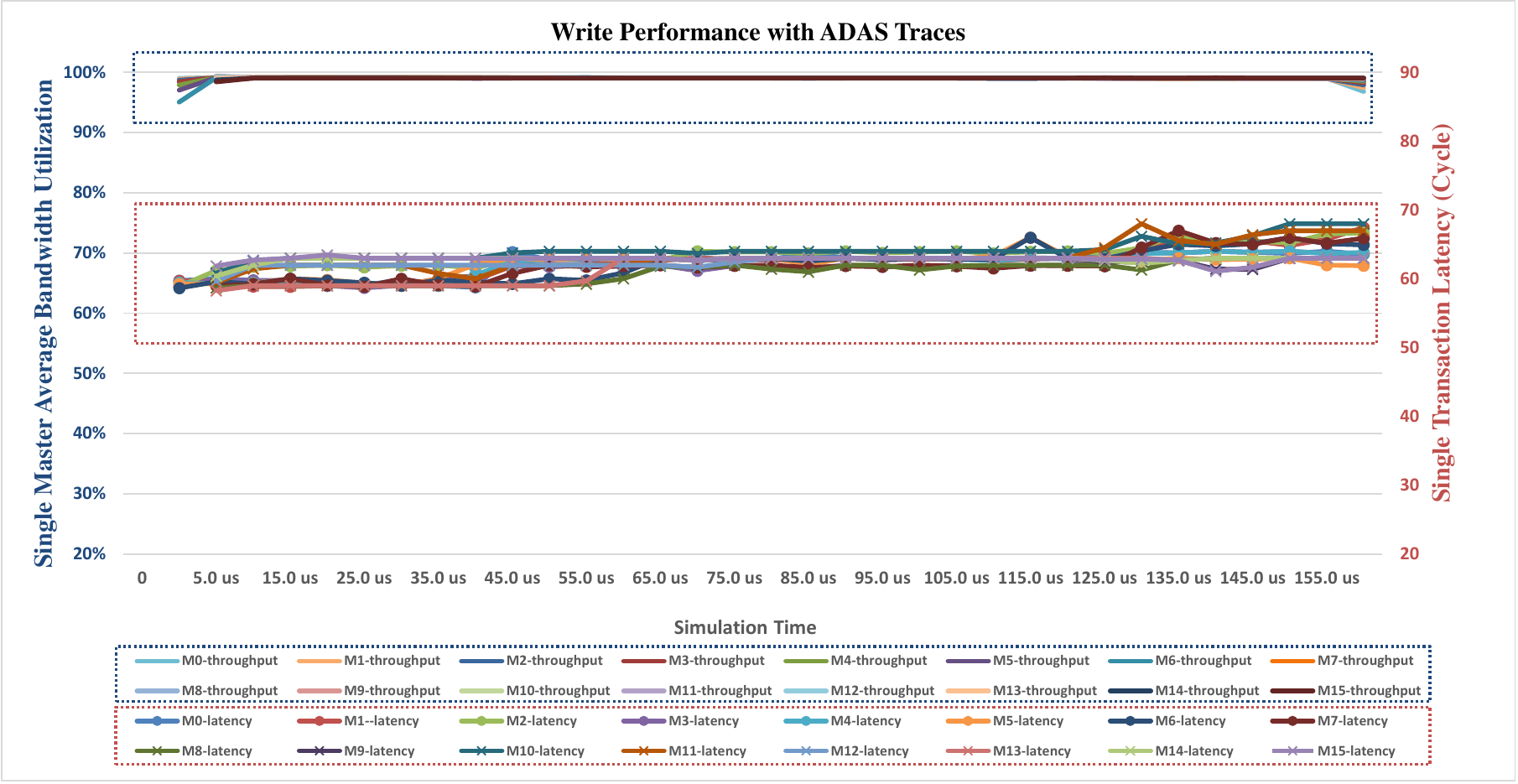}}
\caption{The Write Performance with Different ML and ADAS Traces}
\label{sim-wr-latency}
\end{figure}

As shown in Fig. \ref{sim-rd-wr-latency}, the “Ideal” number is calculated with 100\% data bus utilization. The number of cycles to transfer such an amount of data can be calculated as the size of the transfer payload divided by the size of the data width of the bus interface. For example, it takes $ \frac{4*1024}{32} = 128 $ cycles to transfer a 4 KB data in an ideal condition. 
It clearly shows that after the initial datapath pipeline latency, which is thirty-two cycles for the read, is incurred in the first burst, there isn't any more extra latency introduced. It proves that the actual data throughput is close to 100\% for the bulk data read. The bulk write also has a similar result that close to 100\% utilization rate is achieved after first write is completed. Just as shown in Table \ref{tab1}, a bigger average read latency is observed in Fig.\ref{sim-rd-wr-thput} because the number of outstanding commands is set to 16 per master port, which is slightly bigger than the maximum number of outstanding commands set for the burst 16 traffic by design, to achieve the highest throughput. However, the average read access latency is settled at $36$ cycles once the number of read outstanding command is reduced to one. The average latency of the writes is more consistent compared with that of reads because the write data is always current and the splitting buffer is deep enough to take in enough beats and remove bubbles between the multiple writes.\\
\begin{table}[htbp]
\caption{The Latency with Different Number of Outstanding}
\begin{center}
\begin{tabular} { |p{1cm}||p{2cm}|p{2cm}|p{2cm}| }
\hline
\textbf{Setting} & \textbf{\textit{Number of Read Ports}}& \textbf{\textit{Number of Outstanding/Port}}& \textbf{\textit{Stable Average Read Latency}} \\
\hline
$1$ & $16$ &$16$& $222$\\
$2$ & $16$ &$1$& $36$\\
\hline
\end{tabular}
\label{tab1}
\end{center}
\end{table}
One of the heaviest traces is then picked to further stress the architecture. This is one of the traces obtained from our earlier version of SoC. In this trace, every master has a 2MB memory space reserved for its read and write accesses. Each of the master 0 to 7 runs with memory traffic from an in-house single shot detection network, the size of data ranges from less than 4 KB to around 260 KB; each of the master 8 to 15 reads and writes with ROIs based on a 1080p YUV422 image where each ROI clips at 2 MB if it is bigger. The setup is to mimic that eight PEs run in parallel to process the images from up to eight cameras and all the masters read and write the data per defined in the traces. Fig. \ref{sim-rd-latency} shows the read latencies fluctuate a bit more in ML features and weights than in the image data. This is because a different sub-region of the feature might be read in a different layer's processing. The feature access pattern of repetitively accessing a portion of a line then a jump to the next line leads to more bank access conflicts. The access pattern of the image data is to continuously access across the full ROI one line after another. Also, shorter burst lengths such as burst four or eight are used by the PEs instead of burst sixteen, which is consistently used by the data transfer of image data. These two contribute to the differences in average read latency between the two types of traffic. Other than the above, the overall throughput is still close to 100\%, which is no different from random generated data simulations, so are the latency and throughput for the writes. 
\\
More results obtained from traces with less access intensity are omitted since the overall results are slightly better with almost the same throughput as shown in Fig. \ref{sim-rd-latency} and \ref{sim-wr-latency}. The results correlate well with the architectural analysis that the longer the burst, the better randomization, the easier a consistent access latency and throughput can be achieved.\\
\subsection{Physical Design Results}
\begin{figure}
\centerline{\includegraphics[width=0.7\linewidth, height=6cm]{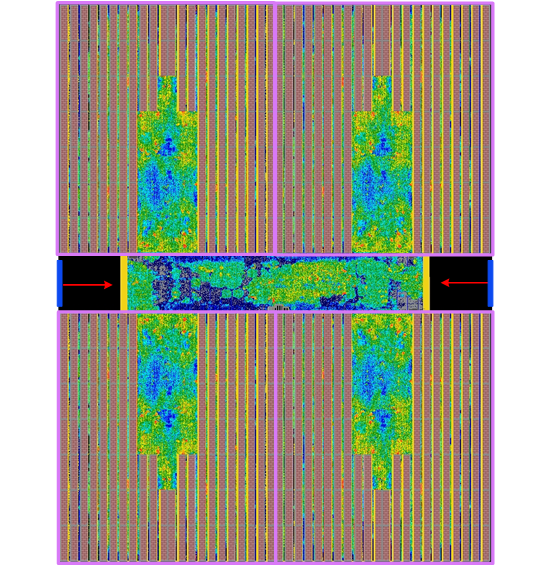}}
\caption{A Snapshot of One Timing Closed Physical Layout}
\label{ssram-layout}
\end{figure}

The architecture has gone through the entire physical design sign-off flow using a TSMC 7nm technology library. As shown in the purple region in Fig. \ref{ssram-layout}, the modularity and hierarchical concept enabled by the architecture makes it very easy to realize such a big design using either Synopsys or Cadence EDA tools. The physical design can take a hierarchical approach that once one cluster is implemented as a hard macro, four clusters are ready to be integrated with proper tie-offs and can be timing closed and signed off in the next higher level even in parallel with a short turnaround time. \\
The NUMA aware architecture enables a much smoother timing closure to otherwise challenging timing paths to memory macros. All the memory macros can run at half speed as that of the interconnect because the randomization of the data return can easily absorb a slightly longer timing path segment to the SRAM macros. This enables us to select SVT (Standard Vt) SRAM macros with higher density and good aspect ratios. The relaxation of the timing on SRAM leads to a 8\% dynamic power reduction and about 11\% leakage reduction for the entire SRAM macros. It also leads to less congestion in layout where one can only see few and tiny red regions (which indicates high congested area) in the timing closed layout shown in Fig. \ref{ssram-layout}. \\

The physical layout is similar to paper layout shown in Fig.\ref{arch-figure} but all the IOs from all accessing masters are split into two groups and arrive from the center openings on the east and west sides. The architecture enables flexible assignment the I/Os to comply top level SoC routing and connection needs. The overall utilization of the silicon area is improved because of the removal I/Os on the north and south sides. One step further, the I/Os can be pushed further toward to the center with two benefits: one is to shorten the timing paths from the I/Os to the memory interconnect; second, it can also spare some areas at the center of the two edges to accommodate some top level SoC routings. The two level of splitting and merging structure strikes a good balance of the performance and wire density for such a wire and memory dominant architecture, the total area comes slightly under 30 mm\textsuperscript{2} with roughly less than 36\% area for the interconnect logic. The utilization inside each cluster is around 40\% and the utilization in the top level is around 35\% for the non-SRAM macro area. This is better than the result reported in \cite{b6}, which low radix switches are used for the interconnect to improve utilization.\\

\section{Conclusion}\label{sec-Conclusion}
We present a new shared memory architecture target to high performance ADAS SoCs. The results show it can provide close to 100\% throughput in both read and write for ML accelerations and ADAS related raw data processing. It realizes consistent QoS for the domain specific payloads and it represents a big and flat memory space with necessary isolation complied to ISO 26262 \cite{b7}, which significantly simplifies the software programming and code management effort. It has a native modular architecture; it can scale up and out with a much smoother and easier implementation flow as demonstrated in section \ref{sec-implementation-Res}. All of the positive results lead to the successful adoptions of this architecture into our production SoCs that target to HAVs. 
\\
Moving forward, we actively investigate to further improving the consistency of the access latency, leveraging the scalable architecture to support even more shared memory and master access ports with 3D integration and Chiplet techniques. \\
\section*{Acknowledgment}

The authors thank the production team of Horizon Robotics to provide relevant data to compare and analyze the benefits of this work. Special thanks to Horizon Robotics' SoC design, functional verification and physical design teams. This work is impossible without their outstanding supports in RTL coding, functional simulation and physical design. \\


\begin{thebibliography}{00}
\bibitem{b1} S. Lin et al., "The Architectural Implications of Autonomous Driving: Constraints and Acceleration". 2018, Proceedings of the Twenty-Third International Conference on Architectural Support for Programming Languages and Operating Systems. Association for Computing Machinery, New York, NY, USA, 751–766. DOI:https://doi.org/10.1145/3173162.3173191
\bibitem{b2} E. Talpes et al., "Compute Solution for Tesla's Full Self-Driving Computer," in IEEE Micro, vol. 40, no. 2, pp. 25-35, 1 March-April 2020, doi: 10.1109/MM.2020.2975764.
\bibitem{b3} V. K. Kukkala, J. Tunnell, S. Pasricha and T. Bradley, "Advanced Driver-Assistance Systems: A Path Toward Autonomous Vehicles," in IEEE Consumer Electronics Magazine, vol. 7, no. 5, pp. 18-25, Sept. 2018, doi: 10.1109/MCE.2018.2828440.
\bibitem{b4} H. Liao, J. Tu, J. Xia and X. Zhou, "DaVinci: A Scalable Architecture for Neural Network Computing," 2019 IEEE Hot Chips 31 Symposium (HCS), 2019, pp. 1-44, doi: 10.1109/HOTCHIPS.2019.8875654.
\bibitem{b5} Nvidia, "NVIDIA Unveils NVIDIA DRIVE Atlan, an AI Data Center on Wheels for Next-Gen Autonomous Vehicles", [Online]. Available: https://nvidianews.nvidia.com/news/nvidia-unveils-nvidia-drive-atlan-an-ai-data-center-on-wheels-fornext-gen-autonomous-vehicles
\bibitem{b6} H. Luan and A. Gatherer, "Combinatorics and Geometry for the Many-ported, Distributed and Shared Memory Architecture," 2020 14th IEEE/ACM International Symposium on Networks-on-Chip (NOCS), 2020, pp. 1-6, doi: 10.1109/NOCS50636.2020.9241708.
\bibitem{b7}ISO26262, [Online], Available: https://www.iso.org/standard/68383.html
\bibitem{b8} B. D. de Dinechin, D. van Amstel, M. Poulhiès and G. Lager, "Time-critical computing on a single-chip massively parallel processor," 2014 Design, Automation \& Test in Europe Conference \& Exhibition (DATE), 2014, pp. 1-6, doi: 10.7873/DATE.2014.110.
\bibitem{b9} Y. Chen et al., "DaDianNao: A Machine-Learning Supercomputer," 2014 47th Annual IEEE/ACM International Symposium on Microarchitecture, 2014, pp. 609-622, doi: 10.1109/MICRO.2014.58.
\bibitem{b10} A. O. Balkan, G. Qu and U. Vishkin, "A Mesh-of-Trees Interconnection Network for Single-Chip Parallel Processing," IEEE 17th International Conference on Application-specific Systems, Architectures and Processors (ASAP'06), 2006, pp. 73-80, doi: 10.1109/ASAP.2006.6.
\bibitem{b11} N. Capodieci, P. Burgio, R. Cavicchioli, I. S. Olmedo, M. Solieri and M. Bertogna, "Real-Time Requirements for ADAS Platforms Featuring Shared Memory Hierarchies," in IEEE Design \& Test, vol. 39, no. 1, pp. 35-41, Feb. 2022, doi: 10.1109/MDAT.2020.3013828.
\bibitem{b12} SAE International, "Taxonomy and Definitions for Terms Related to On-Road Motor Vehicle Automated Driving Systems", [Online], Available: https://www.sae.org/standards/content/
\end{thebibliography}
\end{document}